\documentclass[sigconf, nonacm, screen, 10pt]{acmart}
\usepackage{amsmath}
\usepackage{multirow}
\usepackage{booktabs}
\usepackage{caption}
\usepackage{algorithm}
\usepackage{algorithmicx}
\usepackage{algpseudocode}
\usepackage{listings}
\usepackage{xcolor}

\definecolor{delim}{RGB}{230, 57, 70}          % Red for delimiters
\definecolor{numb}{RGB}{231, 111, 81}         % Orange green for numbers
\definecolor{keycolor}{RGB}{9, 12, 155}       % Dark Blue for keys
\definecolor{valuecolor}{RGB}{60, 185, 137} % Green for values

\lstdefinelanguage{json}{
    columns=flexible,
    numbers=none,                              % Line numbers on the left
    numberstyle=\footnotesize,                        % Small font for line numbers
    frame=none,                                  % Frame around the code
    rulecolor=\color{black},                   % Frame color
    showspaces=false,                          % Do not show spaces
    showstringspaces=false,                    % Do not show spaces within strings
    showtabs=false,                            % Do not show tabs
    breaklines=true,                           % Enable line breaking
    postbreak=\raisebox{0ex}[0ex][0ex]{\ensuremath{\color{gray}\hookrightarrow\space}}, % Symbol at line break
    breakatwhitespace=false,                    % Break at whitespace
    basicstyle=\ttfamily\footnotesize,                % Basic font style
    upquote=true,                              % Use upright quotes
    morestring=[b]",                           % Strings within double quotes
    stringstyle=\color{valuecolor},            % Default color for values
    aboveskip=0pt,                             % Reduce space above
    belowskip=0pt,                             % Reduce space below
    literate=
     *{0}{{{\color{numb}0}}}{1}                % Color numbers
      {1}{{{\color{numb}1}}}{1}
      {2}{{{\color{numb}2}}}{1}
      {3}{{{\color{numb}3}}}{1}
      {4}{{{\color{numb}4}}}{1}
      {5}{{{\color{numb}5}}}{1}
      {6}{{{\color{numb}6}}}{1}
      {7}{{{\color{numb}7}}}{1}
      {8}{{{\color{numb}8}}}{1}
      {9}{{{\color{numb}9}}}{1}
      {:}{{{\textbf{\color{delim}:}}}}{1}               % Color colon
      {,}{{{\textbf{\color{delim},}}}}{1}               % Color comma
      {\{}{{{\textbf{\color{delim}\{}}}}{1}             % Color for opening brace
      {\}}{{{\textbf{\color{delim}\}}}}}{1}             % Color for closing brace
      {[}{{{\textbf{\color{delim}[}}}}{1}               % Color for opening bracket
      {]}{{{\textbf{\color{delim}]}}}}{1}               % Color for closing bracket
      {"capabilities"}{{{\color{keycolor}"capabilities"}}}{14}
      {"data\_format"}{{{\color{keycolor}"data\_format"}}}{13}
      {"id"}{{{\color{keycolor}"id"}}}{4}
      {"last\_sync\_timestamp"}{{{\color{keycolor}"last\_sync\_timestamp"}}}{23}
      {"registration\_timestamp"}{{{\color{keycolor}"registration\_timestamp"}}}{24}
      {"status"}{{{\color{keycolor}"status"}}}{10}
      {"type"}{{{\color{keycolor}"type"}}}{7}
      {"timestamp"}{{{\color{keycolor}"timestamp"}}}{11}
      {"activity\_type"}{{{\color{keycolor}"activity\_type"}}}{14}
      {"details"}{{{\color{keycolor}"details"}}}{9}
      {"device\_id"}{{{\color{keycolor}"device\_id"}}}{10}
      {"old\_version\_id"}{{{\color{keycolor}"old\_version\_id"}}}{10}
      {"new\_version\_id"}{{{\color{keycolor}"new\_version\_id"}}}{10}      {"location"}{{{\color{keycolor}"location"}}}{10}
      {"latitude"}{{{\color{keycolor}"latitude"}}}{10}
      {"longitude"}{{{\color{keycolor}"longitude"}}}{11}
      {"description"}{{{\color{keycolor}"description"}}}{13}
      {"access\_methods"}{{{\color{keycolor}"access\_methods"}}}{16}
      {"api\_endpoint"}{{{\color{keycolor}"api\_endpoint"}}}{14}
      {"protocols"}{{{\color{keycolor}"protocols"}}}{11}
      {"owner"}{{{\color{keycolor}"owner"}}}{7}
}

\AtBeginDocument{%
  }

\usepackage{paralist}
\usepackage{enumitem}
\usepackage{graphicx}
\usepackage{subcaption}
\usepackage{amsmath}

\renewcommand\footnotetextcopyrightpermission[1]{}
\makeatletter
\def\@copyrightspace{\relax}
\makeatother

\begin{document}


\title{The Streetscape Application Services Stack (SASS): Towards a Distributed Sensing Architecture for Urban Applications}

\author{
Navid Salami Pargoo$^{1,*}$, Mahshid Ghasemi$^2$, Shuren Xia$^1$, Mehmet Kerem Turkcan$^2$, Taqiya Ehsan$^1$, Chengbo Zang$^2$, Yuan Sun$^1$, Javad Ghaderi$^2$, Gil Zussman$^2$, Zoran Kostic$^2$, Jorge Ortiz$^{1,*}$
}
\affiliation{
$^1$WINLAB, Rutgers University \state{New Jersey} \country{USA}\\
$^2$Columbia University \state{New York} \country{USA}
}
\email{
Email: {navid.salamipargoo, shuren.xia, taqiya.ehsan, ys820, jorge.ortiz}@rutgers.edu;
}
\email{
{mahshid.ghasemi, mkt2126, cz2678, jg3465, gil.zussman, zk2172}@columbia.edu 
}
\thanks{* Corresponding authors.}

\renewcommand{\shortauthors}{Salami Pargoo et al.}

\begin{abstract}

As urban populations grow, cities are becoming more complex, driving the deployment of interconnected sensing systems to realize the vision of smart cities. These systems aim to improve safety, mobility, and quality of life through applications that integrate diverse sensors with real-time decision-making. Streetscape applications—focusing on challenges like pedestrian safety and adaptive traffic management—depend on managing distributed, heterogeneous sensor data, aligning information across time and space, and enabling real-time processing. These tasks are inherently complex and often difficult to scale. The \textbf{Streetscape Application Services Stack (SASS)} addresses these challenges with three core services: \emph{multimodal data synchronization}, \emph{spatiotemporal data fusion}, and \emph{distributed edge computing}. By structuring these capabilities as clear, composable abstractions with clear semantics, SASS allows developers to scale streetscape applications efficiently while minimizing the complexity of multimodal integration.

We evaluated SASS in two real-world testbed environments: a controlled parking lot and an urban intersection in a major U.S. city. These testbeds allowed us to test SASS under diverse conditions, demonstrating its practical applicability. The Multimodal Data Synchronization service reduced temporal misalignment errors by \emph{88\%}, achieving synchronization accuracy within \emph{50 milliseconds}. Spatiotemporal Data Fusion service improved detection accuracy for pedestrians and vehicles by over \emph{10\%}, leveraging multicamera integration. The Distributed Edge Computing service increased system throughput by \emph{more than an order of magnitude}. Together, these results show how SASS provides the abstractions and performance needed to support real-time, scalable urban applications, bridging the gap between sensing infrastructure and actionable streetscape intelligence.

\end{abstract}

%    <concept>
%        <concept_id>10010520.10010553</concept_id>
%        <concept_desc>Computer systems organization~Embedded and cyber-physical systems</concept_desc>
%        <concept_significance>500</concept_significance>
%        </concept>
%    <concept>
%        <concept_id>10010147.10010257</concept_id>
%        <concept_desc>Computing methodologies~Machine learning</concept_desc>
%        <concept_significance>500</concept_significance>
%        </concept>
%    <concept>
%        <concept_id>10003120.10003138</concept_id>
%        <concept_desc>Human-centered computing~Ubiquitous and mobile computing</concept_desc>
%        <concept_significance>500</concept_significance>
%        </concept>
%  </ccs2012>

% \ccsdesc[500]{Computer systems organization~Embedded and cyber-physical systems}
% \ccsdesc[500]{Computing methodologies~Machine learning}
% \ccsdesc[500]{Human-centered computing~Ubiquitous and mobile computing}

% \keywords{Sensing Framework, Multimodal Sensing, Smart Cities}

%   \includegraphics[width=\textwidth]{figs/architecture.pdf}
%   \caption{The framework architecture.}
%   \label{fig:arch}


\settopmatter{printfolios=true}
\maketitle

\section{INTRODUCTION}

With rapid urbanization, cities are evolving into complex, sensor-rich environments, equipped with a vast array of sensors, actuators, and communication infrastructure \cite{bibri2017smart}. These urban settings are essentially large-scale distributed systems, generating massive data streams that present new opportunities for streetscape applications focused on enhancing safety, accessibility, and overall urban livability \cite{zanella2014internet}. Streetscape applications address high-impact, real-time challenges, such as improving pedestrian safety at intersections—critical given that in the United States alone, intersections were the site of 1,705 pedestrian fatalities in 2022, representing 23\% of all pedestrian deaths that year \cite{iihs_fatality_statistics}. By leveraging advanced sensing and computing capabilities, these applications can respond to the immediate needs of urban environments, making interactions within cities safer and more context-aware. Yet, despite the increasing instrumentation available in smart city testbeds, creating effective urban applications remains challenging due to the lack of a unified framework that offers consistent abstractions and controlled access to shared resources \cite{santana2017software}.

Current urban systems tend to be isolated, vertically integrated stacks that are tightly bound to specific hardware and software configurations. They often fail to integrate diverse data sources, resulting in fragmented information and limited situational awareness. Applications tailored to one environment are difficult to adapt to others, which severely limits scalability and reusability across different urban settings. Additionally, privacy concerns are frequently overlooked, hindering user trust and adoption. These siloed systems often lack high-level programmability and enforceable access controls, making it hard to bridge the gap between application policies and low-level system operations. Without consistent abstractions and robust access controls, developers are left with ad-hoc solutions rather than a dependable, standardized approach to harness the potential of smart city infrastructure \cite{wu2024deep}.

To address these challenges, we introduce the \textbf{Streetscape Application Services Stack (SASS)}, a structured framework that supports the development of complex urban applications. SASS tackles the limitations of previous frameworks by providing a modular, distributed application stack designed specifically for smart cities. With an SDK and a set of APIs, SASS abstracts the complexity of underlying hardware and software configurations, enabling developers to work with diverse urban sensors and components as manageable, composable units.

SASS is built to meet the specific demands of urban applications, which require precise multimodal data synchronization, spatiotemporal data fusion, and low-latency edge computing. It synchronizes data streams from heterogeneous sensor types, integrates spatially distributed data over time, and performs edge-based processing to reduce latency and conserve bandwidth. By offering these core services as modular, composable components, SASS simplifies the creation and deployment of applications in dynamic urban environments, enabling flexible and robust solutions that adapt to varied infrastructure setups.

We evaluated SASS’s performance across different real-world testbeds: a controlled parking lot and a city-scale mobile wireless testbed in New York City (COSMOS) \cite{raychaudhuri2020challenge}. These evaluations confirm SASS’s ability to support scalable, high-performance urban applications by providing the necessary modular components and architectural support.

The key contributions of this work are as follows:
\begin{itemize}[topsep=0pt, leftmargin=*]
    \item \textbf{Introduction of the Streetscape Application Services Stack:} SASS provides a novel, modular framework with standardized abstractions and controlled resource access, addressing the programmability and portability gaps in existing urban systems. Through its SDK and APIs, SASS facilitates application development in complex urban settings.
    
    \item \textbf{Architectural Design for Core Urban Application Properties:} We identify three critical properties for urban applications—multimodal synchronization, spatiotemporal data fusion, and low-latency edge processing—and design specialized services to support them. These services provide a scalable foundation that simplifies data synchronization, integration, and real-time processing across diverse urban sensors.

    \item \textbf{Implementation of Advanced Application Services:} We present three specialized services:
    % , each leveraging components of the SASS framework to support real-time urban sensing and analytics: 
    Multimodal Sensor Synchronization, Multicamera Detection, and Distributed Edge Processing. These services introduce new techniques, including a two-stage event-based synchronization algorithm, a neural network-based multicamera fusion model, and a decay-based dynamic scheduling algorithm.
\end{itemize}

While existing smart city frameworks have tackled aspects of urban sensing and processing, they typically lack the modularity and composable abstractions necessary for handling the complexity of diverse, distributed urban sensors at scale. SASS addresses these limitations by providing a structured, service-oriented framework tailored to the demands of real-time, multimodal streetscape applications. By bridging the gap between low-level sensor management and high-level application logic, SASS enables scalable and adaptable solutions across various urban testbeds. In the following section, we examine prior work on urban sensing frameworks and multimodal data processing, outlining the key challenges that have influenced SASS’s design.

% The rest of this paper is organized as follows. Section \ref{related} provides a review of the literature. Section \ref{framework} introduces our proposed system architecture framework tailored to smart streetscape applications. The methodological details of our algorithms are presented in Section \ref{method}, while Section \ref{experiments} presents the evaluation of their results. Finally, the conclusion is discussed in Section \ref{conclusion}.

\section{RELATED WORK} \label{related}

The Streetscape Application Services Stack (SASS) distinguishes itself as a novel system architecture that integrates multimodal data synchronization, spatiotemporal data fusion, and edge computing, addressing critical gaps in real-time, distributed urban applications. In this section, we contrast SASS capabilities with existing work across three principal application domains.

\subsection{Multimodal Data Synchronization}

In urban environments, effective data synchronization across diverse sensor modalities is essential for applications requiring coordinated interpretation of events. Early efforts include Spinello et al.'s work on pedestrian detection and tracking using 2D and 3D laser range finders and cameras \cite{spinello2010people}, and Piadyk et. al's \textit{StreetAware} dataset \cite{piadyk2023streetaware} which employed multimodal integration for pedestrian tracking, while \textit{GruMon} leveraged smartphones to detect pedestrian clusters \cite{sen2018grumon}. More recently, Sukel et al. advanced audio-visual data fusion for micro-event classification in urban spaces \cite{sukel2019multimodal}. 

SASS builds on these foundations by offering a robust \textit{Data Synchronization Service} that aligns diverse data streams with advanced timestamp alignment and buffering, crucial for high-precision applications such as assistive navigation and emergency response coordination. By providing a \textit{Multimodal Data API}, SASS enables developers to access synchronized data streams seamlessly, meeting the timing requirements of urban intelligence systems.

\subsection{Spatiotemporal Data Fusion}

Spatiotemporal fusion across distributed sensors is pivotal for applications that monitor dynamic urban patterns over time and space. Studies by Brunetti et al. and Tian et al. illustrate the effectiveness of multi-sensor setups for tracking urban pedestrian movement \cite{brunetti2018computer, tian2019pedestrian}. However, limitations in these systems, particularly regarding scalability and cross-location adaptability, remain.

SASS overcomes these challenges by employing a dedicated \textit{Data Fusion Engine} to aggregate sensor data based on proximity and spatial parameters. These components enable continuous, high-resolution pedestrian and vehicle tracking necessary for adaptive traffic optimization and surveillance, allowing SASS to deliver robust monitoring with minimal latency across urban scales.

\subsection{Edge Computing for Real-Time Data Processing}

Edge computing has been explored for smart city applications to reduce latency and improve data efficiency, as demonstrated by Shi et al. and Yu et al. in traffic and event monitoring contexts \cite{shi2016edge, yu2017survey}. Nonetheless, scalability and resource management in edge infrastructures present ongoing technical challenges, particularly for urban deployments that require adaptive, low-latency responses.

SASS extends real-time processing capabilities for smart cities through its \textit{Edge Computing and Distributed Processing Services}, which facilitate localized data analysis for applications such as adaptive signal control. The \textit{Resource Management Layer} further optimizes computational resources across nodes, providing reliable real-time data handling. SASS' efficient \textit{Communication Middleware} ensures high-speed data exchange, making it ideal for large-scale, latency-sensitive applications.

\subsection{Privacy-Preserving Techniques in Urban Applications}

As urban systems handle increasingly sensitive data, privacy-preserving mechanisms are critical. Liu et al. propose a privacy-preserving approach using multi-armed bandits for IoT \cite{liu2020privacy}, while Miao et al. use crowd-sensing techniques to protect user privacy \cite{miao2019privacy}. Federated learning \cite{mcmahan2017communication, guo2021multi} and blockchain \cite{vangala2021blockchain} have also shown promise in decentralized data security.

SASS incorporates a privacy-focused architecture, combining edge-based anonymization and server-side measures, addressing privacy requirements in pedestrian tracking applications. This design allows SASS to meet regulatory privacy standards, which is critical for secure and compliant urban deployments.

\subsection{Pedestrian Safety and Mobility Applications}

There has been extensive research focusing on pedestrian safety, such as \textit{WheelShare} for accessible routing \cite{guo2018wheelshare} and \textit{Ghost-Probe} for blind spot detection \cite{zhang2020ghost}. Barón et al. studied urban walkability factors \cite{baron2018walkability, baron2018investigating}, which impact pedestrian experiences in urban areas.

Through components like the \textit{Multimodal Synchronization Module}, SASS advances applications for real-time safety interventions, making it particularly valuable for pedestrian detection, health monitoring, and emergency response. This integration of safety and mobility features in a single framework represents a significant contribution to urban intelligence.

\subsection{IoT and Big Sensor Data Systems for Smart Cities}

General IoT frameworks, including \textit{MACeIP} and \textit{REIP}, have enhanced urban system management \cite{maceip, reip}, while the Smart City Framework promotes interoperability in transportation \cite{scf}. SASS surpasses traditional systems by focusing on smart streetscape with an emphasis on real-time pedestrian and traffic monitoring. Its modular architecture allows tailored, low-latency analytics for applications like adaptive traffic control and emergency response.

\section{FRAMEWORK DESIGN} \label{framework}
To clarify the design of the Streetscape Application Services Stack (SASS), we introduce three real-world applications that emerged from smart city testbed initiatives. These applications highlight the specific needs and challenges of streetscape applications, framing the requirements that SASS is designed to meet.

\subsection{Motivating Applications} \label{motivating}
    \textit{Waypoint Finding for Navigation Application}: This application assists visually impaired individuals in navigating urban areas by providing real-time guidance. It combines data from multiple sources, such as GPS, cameras, and wearable devices, to offer accurate directions and obstacle avoidance. The application relies on multimodal data synchronization to ensure guidance is both accurate and responsive to the user’s immediate surroundings~\cite{jain2024streetnav}. 
    
    \textit{Adaptive Traffic Signal for Extended Cross Time Application}: This application aims to improve pedestrian safety by dynamically adjusting traffic signal timings to accommodate individuals with varying mobility needs, like the elderly or those with disabilities. It depends on spatiotemporal data fusion from cameras and wearables to detect pedestrians and estimate crossing times, requiring low-latency processing at the edge to adjust signals in real-time~\cite{fu2023federated, mo2022cvlight}.
    
    \textit{Pedestrian and Vehicle Detection for Urban Analytics Application:} This application monitors pedestrian and vehicle flows across multiple intersections to support urban planning and safety analysis. It integrates data from distributed sensors, requiring spatiotemporal fusion to accurately track movement. Edge computing is used to handle high data volumes and enable real-time analytics~\cite{ghasemi2021auto,ghasemi2022real,sun2024optimizing,wu2023multi}.

Each of these applications relies on integrating and processing data from diverse, distributed sensors in real-time to deliver hyper-local, context-aware services. \emph{They all require the ability to synchronize multimodal data, fuse spatiotemporal information, and process data at the edge to meet performance requirements.}

\subsection{Architectural Implications} \label{sec:arch_imp}
Our experience developing applications on COSMOS testbed highlighted the need for shared services and higher-level abstractions to streamline development and reduce repeated work. Building each application independently revealed recurring challenges, such as managing heterogeneous data, synchronizing data streams, and ensuring low-latency processing. These challenges, which we observed across urban applications in the literature, are foundational for streetscape environments~\cite{santana2017software,wu2024deep,al2015applications}.

From this analysis, we identified three properties that are essential to address the multimodal, distributed nature of urban environments:

\begin{itemize}[leftmargin=*]
    \item \textbf{Multimodal Data Synchronization:} Synchronizing data from different sensor types in time is essential for coherent analysis and processing. Without proper alignment, data fusion can produce misleading or incorrect results. Applications that rely on this property include assistive navigation for visually impaired users \cite{jain2024streetnav} and emergency response systems that need precise, real-time situational data~\cite{olivier2022data,olivier2023bayesian}.

    \item \textbf{Spatiotemporal Data Fusion:} Integrating data across time and spatial locations is essential for capturing and analyzing dynamic events and patterns in the urban landscape. Applications like multi-camera pedestrian tracking \cite{lima2021generalizable} and traffic flow optimization \cite{tang2019cityflow} depend on this capability to interpret movement and detect changes across different areas.

    \item \textbf{Edge Computing and Distributed Processing:} Processing data close to the source reduces latency and lowers bandwidth usage, which is needed to support real-time applications. Distributed processing improves scalability by balancing computation across multiple devices. This property is central to real-time analytics in traffic management systems \cite{ghasemi2022real} and adaptive traffic signal control \cite{fu2023federated}.

\end{itemize}
        
These core properties informed SASS’s architectural design. To support scalable, efficient urban applications, SASS provides services and abstractions that address these common needs. By focusing on multimodal data synchronization, spatiotemporal data fusion, and edge computing, SASS enables developers to build applications that integrate data seamlessly, perform real-time analysis, and leverage edge resources, while reducing the inherent complexity of streetscape development.

\section{SYSTEM ARCHITECTURE DESIGN AND IMPLEMENTATION} \label{sec:design}
SASS is a modular framework designed to handle the unique demands of urban applications by providing scalable tools for development, deployment, and management. Structured into distinct layers and services, SASS abstracts core complexities and offers comprehensive support for building urban data-driven applications.

\begin{figure*}
    \centering
    \includegraphics[width=0.9\textwidth]{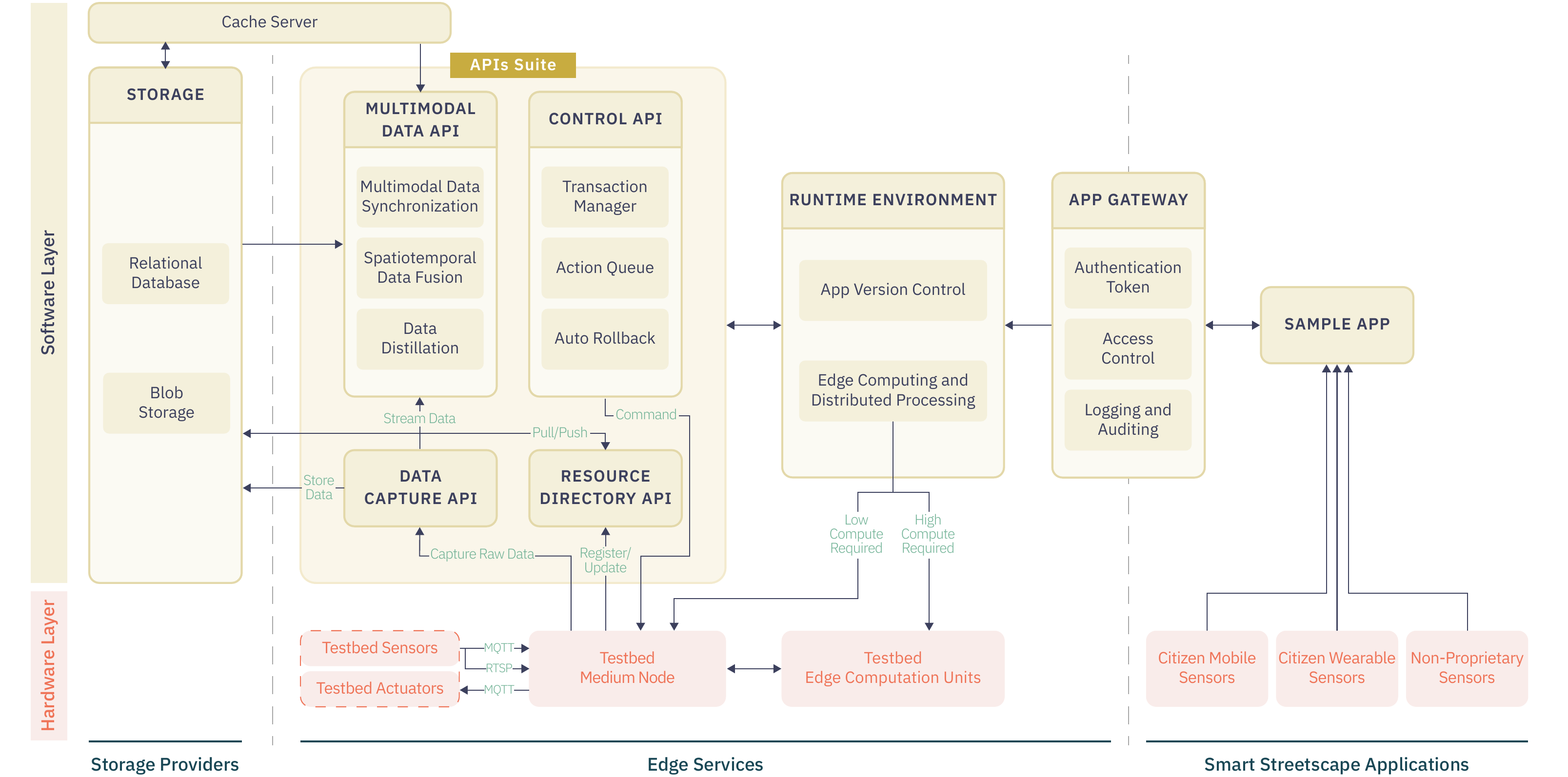}
    \caption{SASS system architecture with core services for multimodal data synchronization, spatiotemporal fusion, and edge computing. Key subsystems support device management, data routing, and distillation, while the Control API ensures transaction reliability with rollback and action queues. The App Gateway secures access through authentication and auditing, and the Runtime Environment manages resources across edge nodes and various sensor types.}
    \label{fig: system}
\end{figure*}

\subsection{Overall System Architecture} \label{sec:arch}
The SASS architecture consists of six main parts (see Figure~\ref{fig: system}): (1) Hardware Layer, (2) API Suite, (3) Data Storage, (4) Runtime Environment, (5) APP Gateway, and (6) Smart Streetscape Applications. Each layer has a defined role that contributes to efficient, scalable, and secure data processing and application support.

The \textbf{Edge Hardware Layer} acts as the system’s base, comprising testbed sensors, actuators, and edge nodes for localized data capture and command execution. Sensors and actuators communicate with medium nodes over MQTT and RTSP to relay raw data and receive commands, and medium nodes further link these devices with software services. Edge nodes execute high-compute tasks and process data near its source to reduce latency. This setup also includes mobile and wearable citizen sensors, such as IMUs and GPS, adding diverse data to the system without being restricted to proprietary infrastructure.

\textbf{Data Storage} manages both structured and unstructured data, supporting relational databases for structured sensor data and configurations, and blob storage for unstructured data such as videos and images. A cache server temporarily stores frequently accessed data to enhance I/O performance.

The \textbf{API Suite} in SASS consists of four key APIs that enable seamless interaction between applications and the system's underlying services. The \textbf{Multimodal Data API} is designed to synchronize and fuse data from a variety of sources, providing coherent and aligned inputs for applications. It also ensures privacy compliance by applying access-based data distillation mechanisms. This API is particularly critical for applications relying on diverse sensor modalities to deliver accurate and integrated outputs.

The \textbf{Control API} facilitates actuator interactions, ensuring transaction integrity through an action queue and rollback mechanisms. By managing command consistency and recovering from partial failures, this API maintains the stability and reliability of actuator operations, enabling robust system control.

The \textbf{Data Capture API} supports the acquisition and storage of both raw and processed data from various sensors and applications. It streams collected data into a centralized database, simplifying its integration into multimodal fusion workflows and analytical pipelines, which enhances system-wide data accessibility.

Finally, the \textbf{Resource Directory API} manages the registration and monitoring of IoT devices. It provides developers and users with a unified interface to track device status and efficiently manage resources. This API enhances operational oversight and ensures the effective deployment of connected devices across the system.

The \textbf{Runtime Environment} on edge nodes integrates with Git for app version control, ensuring nodes stay synchronized with the latest versions and allowing quick rollbacks if issues arise. It provides the necessary infrastructure for edge computing and distributed processing, handling resource management and network coordination to streamline task execution.

The \textbf{App Gateway} secures access to SASS’s services, controlling entry points with authentication and access control. By segmenting the runtime environment, the Gateway prevents unauthorized data access and maintains user privacy. Comprehensive logging and auditing support traceability and system monitoring.

At the top, \textbf{Smart Streetscape Applications} make use of SASS services and data to address urban challenges. Applications can directly interact with citizens or support city agencies by securely sharing data for urban planning and public safety.

\subsection{Core Services} \label{sec:fund}
\textit{SASS Core Services} is the backbone for managing IoT devices in urban settings, with a modular framework that facilitates device registration, secure access, and real-time data flow management.

\begin{figure*}[th]
    \centering
    \begin{subfigure}[t]{0.46\textwidth}
        \centering
        \begin{lstlisting}[language=json]    
{"device_id":"camera-001","type":"sensor","location":{"latitude":40.00,"longitude":-70.00,"description":"Alpha St."},"capabilities":["video_stream","detect","track"],"data_format":"H.264","access_methods":{"api_endpoint":"https://generic-endpoint.org/","protocols":"RTSP"},"status":"online","last_sync_timestamp":"2024-11-07T13:24:02.0923","registration_timestamp":"2024-08-05T11:19:31.5754","owner":"Testbed"}
        \end{lstlisting}
        \caption{Device Registration Information }
        \label{fig:device_list}
    \end{subfigure}
    \hfill
    \begin{subfigure}[t]{0.28\textwidth}
        \centering
        \begin{lstlisting}[language=json]    
{"timestamp":"2024-11-07T13:49:53.6500","activity_type":"rollback","details":{"device_id":"sensor-001","old_version_id":"e4a7e88b-f32b4cf5-fca31f3f","new_version_id":"c0aa938e-0c8d4c39-ca4c6681"}}
    \end{lstlisting}
        \caption{Rollback Log (Version Control)}
        \label{fig:rollback}
    \end{subfigure}
    \hfill
    \begin{subfigure}[t]{0.25\textwidth}
        \centering
        \begin{lstlisting}[language=json]
{"timestamp":"2024-10-05T21:19:45.3932","activity_type":"update","details":{"device_id":"sensor-001","status":"maintenance","last_sync_timestamp":"2024-10-05T21:19:45.388"}}
        \end{lstlisting}
        \caption{Device Status Log}
        \label{fig:device_log}
    \end{subfigure}
    \hfill
    \caption{Illustrative examples of the JSON-based data format used in SASS for device management and monitoring.}
    \label{XXX}
\end{figure*}

\subsubsection{Key Components and Methodology}
Core Services consists of several key components that together ensure secure and efficient urban IoT management.

The \textbf{User Management and Access Control} component manages access through a role-based system using JSON Web Tokens (JWT). Each token encodes user roles and permissions, enforcing access control across all requests. Permissions are assigned based on user roles and validated at each request to secure sensitive data.

The \textbf{Resource Directory} registers and monitors devices, storing attributes like device type, location, and access methods. Using SQLAlchemy ORM, a polymorphic base class manages diverse device types, allowing easy extension for future devices. Device-specific tokens authenticate status updates, while version histories provide tracking and operational visibility.

The \textbf{Data Capture} component acquires data from sensors, performing initial filtering, cleaning, and compression. Data is tagged with metadata such as location and timestamps, then directed to the Multimodal Data API for further processing or stored in SASS’s database.

A \textbf{Data Distillation} pipeline balances data utility and privacy using anonymization, aggregation, and temporal delay techniques. Anonymization uses cryptographic hashing and location generalization, aggregation provides summary statistics, and temporal delay prevents real-time tracking by adding a time buffer. Future work includes exploring differential privacy and federated learning to enhance privacy protections.

The \textbf{Logging and Auditing} component tracks system activities like device registrations and data access, storing logs in JSON format with categorized timestamps and event types. This logging framework supports troubleshooting and compliance by providing a complete interaction history.

\textbf{Version Control and Rollback} leverages a UUID-based snapshot system that simplifies rollbacks by storing each configuration as a complete state, ensuring that administrators can easily restore previous configurations if needed.

The \textbf{Transaction Manager} acts as an intermediary between applications and actuators, ensuring control commands are accurately executed. The \textbf{Action Queue} sequences commands to prevent conflicts and includes an auto-rollback mechanism for error handling, restoring stable configurations when necessary. Integration testing ensures reliable operation across system components.

\subsubsection{System Workflow}
Core Services handle device registration, data flow, control operations, authentication, configuration management, and monitoring through a series of streamlined processes.

Applications interact with SASS through JWT-based authentication, obtaining temporary tokens that encode access permissions. When a device registers, it submits attributes like location and capabilities (see Figure \ref{fig:device_list}), and after validation, the system generates a device token for secure communication. Device status and configuration changes are logged, with rollback versions stored in the version control system. IoT devices transmit data and accept control commands through secure channels.

For data flow, devices send status updates and data through secure channels authenticated by device tokens. Data Capture manages incoming data, tagging it with metadata, synchronizing streams, and storing data in the system. For control operations, applications initiate commands through the Transaction Manager, which translates commands into actuator-friendly formats, sequences them in the Action Queue, and handles auto-rollback if errors occur (see Figure \ref{fig:rollback}). Control activities are logged, providing traceability (see Figure \ref{fig:device_log}).

\begin{figure*}[th]
    \centering
    \includegraphics[width=\textwidth]{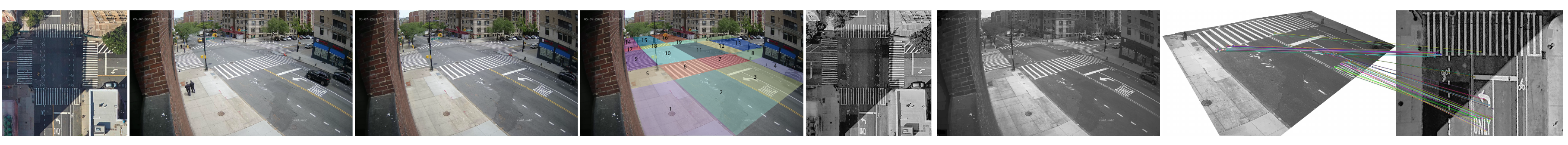}
    \caption{Perspective transformation workflow using satellite and camera imagery. Left to right: (1) Original satellite image, (2) Original camera image, (3) Extracted background, (4) Area segmentation, (5) Histogram-equalized satellite image, (6) Histogram-equalized camera image, (7) Feature matching between satellite and camera images.}
    \label{fig:transformation}
\end{figure*}

\subsection{Multimodal Data Synchronization Services} \label{sec:synch}
The \textit{Multimodal Data Synchronization Service} aligns streams from sensors with varying rates, formats, and time accuracies, ensuring consistency across data sources, especially in real-time applications.

Incoming \textit{edge sensor data} streams are timestamped using NTP to maintain accuracy across devices, with clock drift corrected using a Kalman filter. Timestamp corrections are calculated as:
$
T_{\text{corrected}} = T_{\text{local}} + \Delta T_{\text{offset}} + \Delta T_{\text{drift}} \cdot \Delta t
$,
where \( T_{\text{local}} \) is the device's local timestamp, \( \Delta T_{\text{offset}} \) is the fixed time offset between clocks, \( \Delta T_{\text{drift}} \) is the drift rate, and \( \Delta t \) is the elapsed time since the last synchronization.
For \textit{mobile and wearable data}, SASS supports both on-device NTP synchronization and alternative methods to reduce battery and bandwidth demands. Predefined and custom synchronization algorithms, such as Dynamic Time Warping (DTW), allow developers to tailor alignment methods to specific applications.

Adaptive buffering manages inconsistent latencies by adjusting buffer sizes based on observed data arrival times:
$
B = \max(B_{\text{min}}, \beta \cdot \sigma_{\text{arrival}})
$
where \( B_{\text{min}} \) is the minimum buffer time, \( \beta \) is a scaling factor, and \( \sigma_{\text{arrival}} \) is the standard deviation of packet intervals.

\subsection{Spatiotemporal Data Fusion Services} \label{sec:fuse}
The \textit{Spatiotemporal Data Fusion Service} integrates data from spatially distributed sensors, aligning data temporally and spatially without detailed calibration.

\begin{algorithm}[t]
\caption{Decay-Based Dynamic Scheduling Algorithm} \label{alg:decay}
\begin{algorithmic}[1]
\Require Task Queue \( \{T_1, T_2, \ldots, T_n\} \), Decay factor \( \alpha \), Initial priority \( P_{\text{initial}}(T) \)
\State Initialize \( P(T) = P_{\text{initial}}(T) \) and \( W(T) = 0 \) for each \( T \) in Task Queue
\While{true}
    \For{each \( T \) in Task Queue}
        \State \( W(T) \gets \text{current time} - \text{entry time of } T \)
        \State \( P(T) \gets P_{\text{initial}}(T) + \alpha \cdot \log(1 + W(T)) \)
    \EndFor
    \State Sort Task Queue by \( P(T) \) (descending)
    \For{each \( T \) in Task Queue}
        \If{resources available}
            \State Execute \( T \)
            \State Remove \( T \) from Task Queue
        \EndIf
    \EndFor
    \State Wait for the next cycle
\EndWhile
\end{algorithmic}
\end{algorithm}

The \textit{Fusion Workflow Engine} orchestrates data flow, mapping incoming data to a shared spatial and temporal framework, operating fusion at raw, feature, and decision levels. For visual sensors, \textit{Automatic Inverse Perspective Mapping} aligns camera views with satellite imagery by computing a homography transformation matrix \(\mathbf{T}\) (Figure~\ref{fig:transformation}). Using Affine Scale-Invariant Feature Transform (ASIFT) and RANSAC, SASS achieves accurate spatial alignment, even in complex environments. Preprocessing includes background extraction, histogram equalization, and area segmentation for enhanced alignment.
Fused data is tagged with fusion metadata, enabling downstream analysis and storage.

\subsection{Edge Computing and Distributed Processing Services}\label{sec:edge}
The Edge Computing and Distributed Processing service balances computational demands across nodes for efficiency.
Tasks are broken into subtasks and classified by computational requirements. Low-compute tasks are routed to medium nodes for low latency, while high-compute tasks, like 3D pose detection, go to edge computation units. If medium nodes are overloaded, tasks are redirected to high-capacity units, maintaining uninterrupted processing.

Tasks on medium nodes are managed with our proposed \textbf{Decay-Based Dynamic Scheduling algorithm} (Algorithm~\ref{alg:decay}), which prioritizes based on urgency and wait time:
$
P(T) = P_{\text{initial}}(T) + \alpha \cdot \log(1 + W(T))
$
where \( P(T) \) is task priority, \( \alpha \) is a decay factor, and \( W(T) \) is wait time. This approach balances rapid prioritization with gradual growth to prevent low-priority tasks from overtaking high-priority ones, ensuring fairness and responsiveness.
A centralized resource monitor tracks utilization across all nodes, providing real-time data for efficient load balancing. Computation nodes process tasks in batches, optimized for high-throughput environments where immediate response is not critical.

\section{APPLICATIONS METHODOLOGY} \label{method}

SASS enables a range of urban streetscape applications by streamlining data integration and reducing complexity across multimodal inputs. In Section \ref{sec:arch_imp}, we outlined the core properties needed to support these applications, and Sections~\ref{sec:synch}--\ref{sec:edge} detail the modular components of SASS that encapsulate these properties as scalable services. To validate SASS’s capabilities, we present three real-time application services developed on SASS, each demonstrating how the stack handles the integration and processing demands unique to urban streetscapes. This section covers the methodology for each application. All experiments are conducted in a controlled parking lot and COSMOS testbed and are approved by institutional review boards (IRB), maintaining ethical standards.

\subsection{Multimodal Data Synchronization} \label{sensor_synch}

Synchronizing data from different sensor types in urban environments presents a substantial challenge given the diversity of data sources. Using SASS’s Multimodal Data Synchronization Service, we developed a synchronization application that aligns data streams from various sensors with high efficiency. This application highlights how SASS abstracts and simplifies complex synchronization tasks, enabling consistent real-time integration without burdening the developer with low-level implementation details.

\begin{figure}[t]
    \centering
    \includegraphics[width=0.99\linewidth]{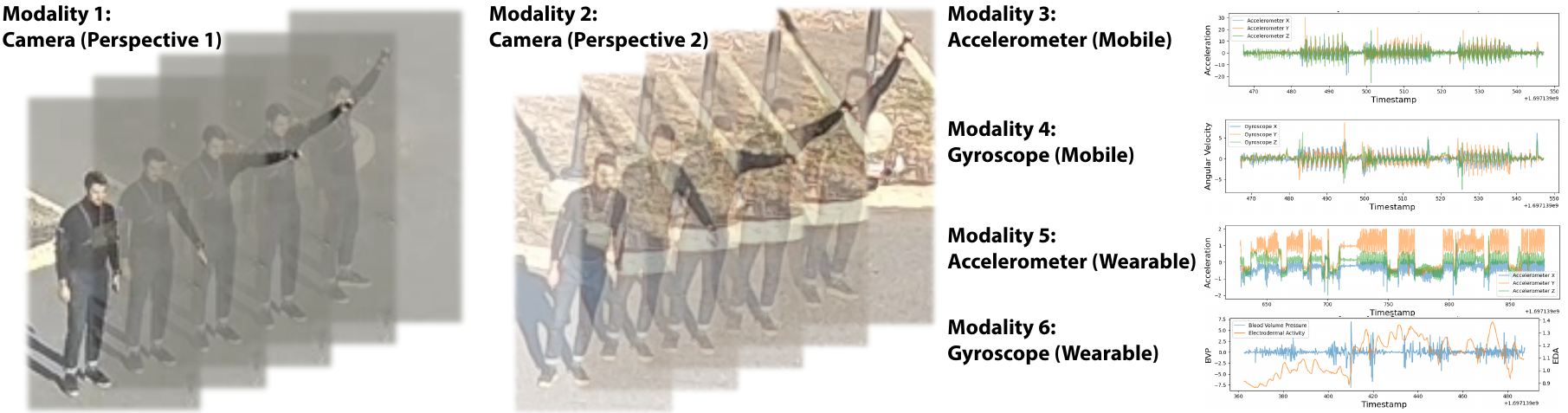}
    \caption{Multimodal Data Streams from various devices to be Synchronized based on Event Detection.}
    \label{fig: sync_data}
\end{figure}

The synchronization setup included two cameras positioned around a parking lot, mobile IMU sensors, and wearable physiological sensors. Each sensor produced temporally linked data streams, as illustrated in Figure \ref{fig: sync_data}.

\subsubsection{Synchronization Algorithm}

We implemented a two-stage algorithm for optimal data alignment across multimodal streams. In the \textit{Coarse Synchronization} stage, we detect predefined events across sensors to establish initial alignment. In \textit{Fine-Tuning}, we minimize any remaining temporal misalignment between modalities.

\subsubsection{Event-Based Synchronization}

Our event for synchronization is a simple hand-raising gesture, detectable across camera feeds and IMU data streams. This simultaneous gesture, with consistent timestamps across sensors, provides a reliable synchronization point.

{\textbf{Gesture Detection in Video Streams}}
Hand gestures in video streams are detected using a combination of 3D pose estimation and Dynamic Time Warping (DTW). The 3D pose estimation identifies keypoints of the hand, tracking its z-coordinate across frames. We then apply DTW-based Barycenter Averaging (DBA) to create a gesture template, which maintains key temporal dynamics \cite{Petitjean2011-DBA, Petitjean2014-ICDM-2}.

To detect matching gestures in new video input, we slide the gesture template over the z-coordinate time series of the hand movement. Detection is confirmed if the DTW distance falls below a preset threshold:
$
    \text{DTW}(S, T) = \min \sum_{(i, j) \in \text{path}} d(s_i, t_j)
$
where \(S\) and \(T\) represent the input signal and template, respectively.

{\textbf{Gesture Detection in IMU Streams}}
For IMU data, we apply Hidden Markov Models (HMM) to identify hand gesture events. IMU data is segmented into sliding windows, extracting features such as mean, variance, SMA, and entropy from accelerometer and gyroscope readings. The HMM model is trained to maximize \( P(\mathcal{O}|\lambda) \), the probability of observing the feature sequence \( \mathcal{O} \) given model \( \lambda \). The Viterbi algorithm decodes the hidden state sequence to identify the gesture within each window.

\subsubsection{Temporal Alignment}

Once detected events are time-stamped in both video and IMU data streams, we align them within a tolerance of ±500 ms. If events fall within this window, they are considered synchronized.

\subsubsection{Fine-Tuning Synchronization}

The fine-tuning stage further refines alignment by pinpointing the start of the hand-drop gesture. The steps include:

\begin{enumerate}[leftmargin=16pt, itemsep=0pt, topsep=4pt]
    \item Applying a Butterworth low-pass filter to reduce noise.
    \item Calculating entropy in a sliding window to detect primary movement patterns.
    \item Identifying entropy peaks to locate the hand-drop start point.
    \item Backtracking to the initial rise in entropy to mark the event’s initiation.
\end{enumerate}

We use the visual stream’s reference timestamp to synchronize the other timestamps, ensuring all streams align precisely.

\subsection{Spatiotemporal Data Fusion}\label{sec:multi-camera-detection}

Spatiotemporal data fusion creates a unified view of urban intersections by combining data from spatially distributed sensors. In environments with overlapping camera coverage, fusion improves object detection and situational awareness. This application service integrates detections from multiple cameras to enhance accuracy and coverage, transforming coordinates into a unified top-down view for real-time scene interpretation. Our test setup included two high-resolution cameras overlooking an intersection from perpendicular angles, capturing video at 3840x2160 resolution.

\begin{figure}[t]
\centering
\includegraphics[width=0.95\textwidth, width=0.95\linewidth]{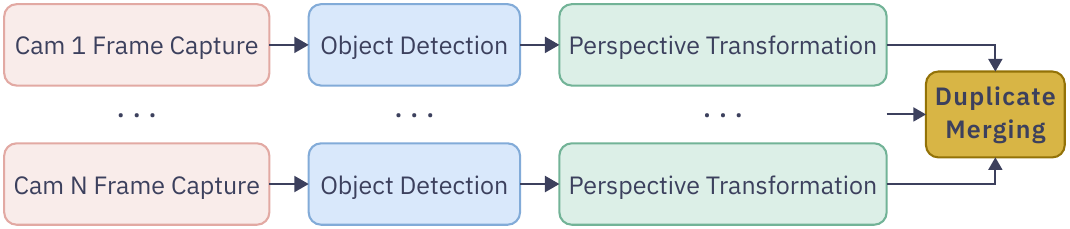}
\caption{Multi-camera integration process. First, an object detection model is applied to each camera. Then, \textit{CoordinateTransformNet} projects bounding boxes to a top-level view, using a Euclidean distance threshold to remove duplicates.}
\label{fig:transformation_process_figure}
\end{figure}

\subsubsection{Fusion Algorithm}
\begin{figure}[t]
    \centering
    \includegraphics[width=1\linewidth]{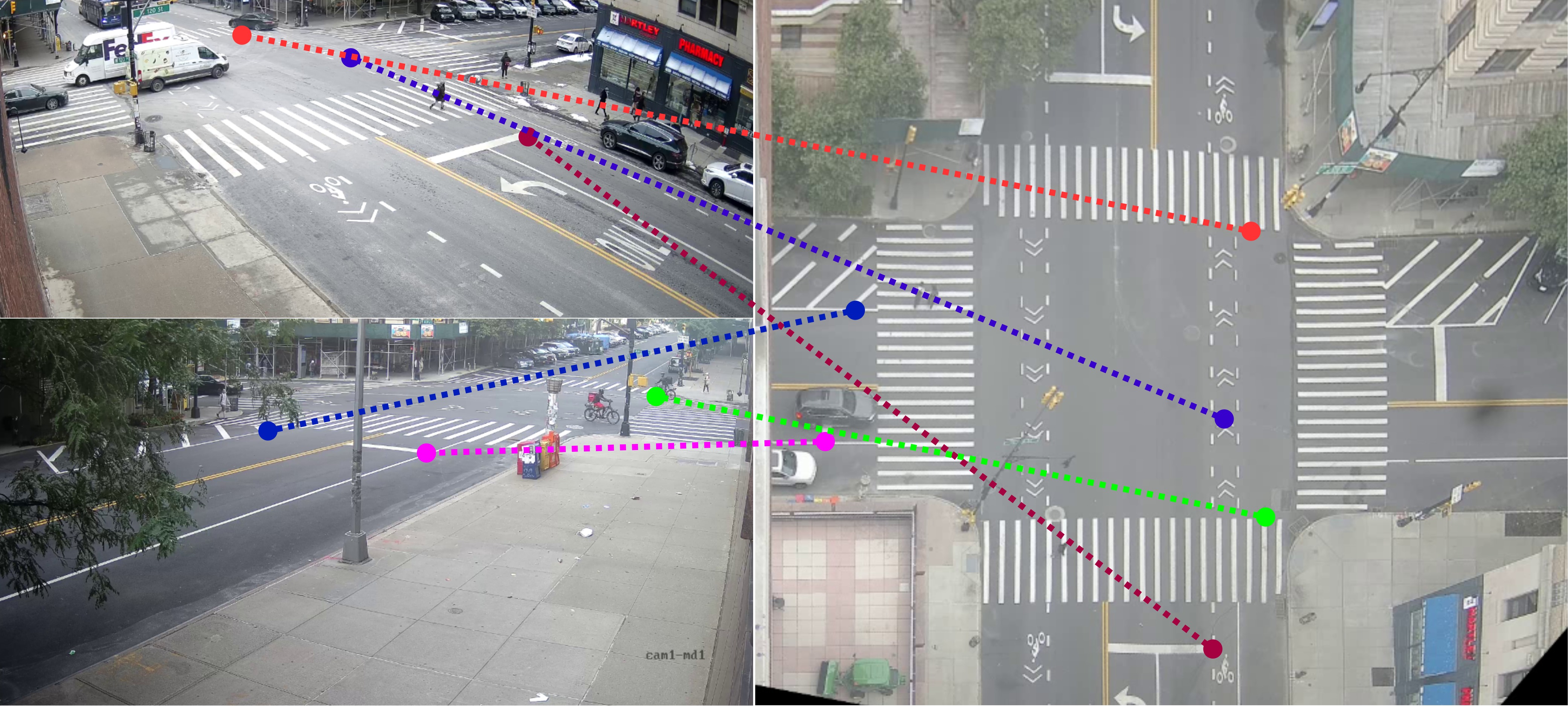}
    \caption{A set of corresponding points from the street-level cameras' perspective and the top-level view perspective is used to train \textit{CoordinateTransformNet}.}
    \label{fig:mapping-points}
\end{figure}

The fusion process involves several sequential steps, shown in Figure~\ref{fig:transformation_process_figure}. We begin by applying YOLOv9e~\cite{wang2024yolov9} to detect pedestrians and vehicles in each video stream. For high-resolution input, slicing-aided hyper inference~\cite{akyon2022sahi} reduces computational overhead. Bounding boxes and class probabilities are generated in each camera’s coordinate system.
Detected coordinates are then transformed using \textit{CoordinateTransformNet}, a lightweight neural network trained using a set of point pairs to map each perspective into a unified top-down view (Figure~\ref{fig:mapping-points}) 
To further refine detections, we apply a Euclidean distance threshold for de-duplication. If the distance between detections from different cameras falls below this threshold, they are treated as duplicates, and we use confidence-weighted averaging to maintain data accuracy.

\subsection{Distributed Edge Computing for Real-Time Data Processing}

Edge processing is necessary for low-latency applications in urban environments. This application service uses SASS’s Edge Computing and Distributed Processing Services to process data at the network edge with minimal latency.
The objective is a real-time processing pipeline that detects and tracks pedestrians, estimates trajectories and poses, and visualizes results on a dashboard. Our setup included mobile IMU sensors and a high-resolution camera in two testbed environments: the controlled parking lot and a dynamic intersection in COSMOS testbed.

IMU readings and camera frames are captured and processed in parallel. For mobile sensor data, only IMU readings from individuals within the target area are used, determined by location metadata.
Frames from edge cameras undergo object detection via YOLOv8~\cite{terven2023comprehensive}. We enhance tracking continuity using the OC-SORT algorithm with shadow tracking for unmatched tracks~\cite{cao2023observation}. Trajectories, 3D bounding box, and 2D poses are estimated, with results displayed on a dashboard for spatial analysis.
Bounding boxes and 2D poses are then fed to MotionBERT~\cite{zhu2023motionbert} for 3D pose estimation, with a Kalman filter applied to smooth keypoint trajectories. Processed data streams are visualized on an interactive dashboard.

%%%%%%%%%%%%%%%%%%%%%%%%%%%%%%%%%%%%%%%%
\section{IMPLEMENTATION AND EVALUATION} \label{experiments}
In this section, we present the implementation of the mentioned application services developed using SASS. We further evaluate each service based on specific performance metrics demonstrating how SASS enables robust, high-performance urban applications across diverse, sensor-rich environments.

\subsection{Multimodal Sensor Synchronization}

The development of our synchronization application service was significantly facilitated by the Multimodal Data Synchronization Services provided by SASS. By leveraging these services, we abstracted away the complexities associated with precise timestamping, clock synchronization, data buffering, and alignment across heterogeneous sensor modalities. This allowed us to focus on high-level synchronization logic and algorithm development, resulting in an efficient and robust synchronization method integrated within the SASS architecture.

The Multimodal Data Synchronization Services within SASS offered core functionalities that streamlined the development process. SASS ensured consistent time references across edge sensors by implementing \textit{precise timestamping mechanisms and clock synchronization protocols}. The services also applied clock drift correction to adjust local timestamps. The services managed variations in data arrival times and sampling rates through \textit{adaptive buffering and resampling mechanisms}. By dynamically adjusting buffer sizes based on real-time monitoring of data arrival variability, the system accommodated network delays and processing latencies without significant impact on synchronization accuracy. SASS provided a \textit{library of predefined algorithms} for data synchronization, including DTW and HMM, which allowed us to utilize these algorithms directly without implementing them from scratch. By utilizing these components, we efficiently implemented our novel, user-defined two-stage synchronization algorithm outlined in Section \ref{sensor_synch}. SASS simplified the development by handling low-level data management tasks, providing reusable algorithms, supporting scalability for the addition of more sensors and modalities, and improving performance by adaptively buffering the data stream.

To further evaluate the effectiveness of our synchronization method, we conducted experiments in our controlled parking lot environment, collecting over an hour of data from various sensors. Eight individuals performed walking and hand-raising/dropping gestures while holding a mobile phone and wearing a wristband. We first evaluated the performance of our gesture detection algorithms in both the visual and IMU data streams.

\begin{figure}
    \centering
    \includegraphics[width=\columnwidth]{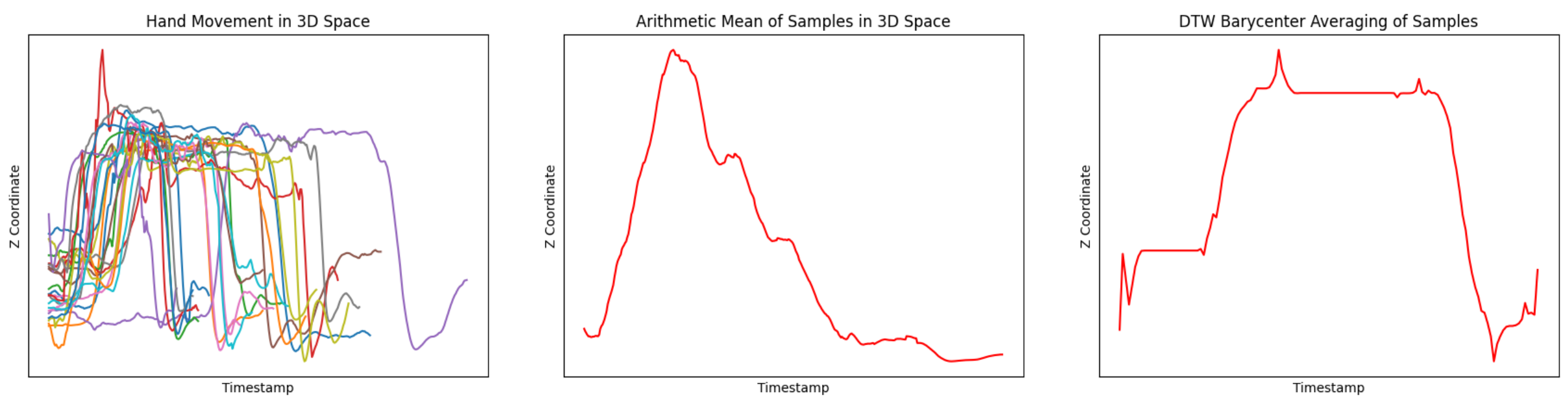}
    \caption{Gesture Samples (Left), their Arithmetic Average w/o Temporal Variations (Middle), and the DTW Barycenter Average w/ Temporal Dynamics used as the Template (Right)}
    \label{fig: dba}
\end{figure}

In the vision domain, a gesture template was created using DTW-based Barycenter Averaging (DBA), which maintained the intrinsic temporal dynamics without distorting key trends (Figure \ref{fig: dba}). The DTW distance threshold was set to 0.8, and the algorithm was applied using sliding windows of 4 seconds. In the IMU domain (mobile devices and wristbands), Features such as mean, variance, standard deviation, Signal Magnitude Area (SMA), and entropy were extracted from the accelerometer and gyroscope signals. The HMM was trained over multiple iterations to maximize the likelihood of observing the feature sequence given the model. The precision, recall and F1-score were calculated for both domains and reported in Table~\ref{table: accuracy}.

After detecting synchronization events in both modalities, we applied our synchronization method to align the data streams. We quantified the effectiveness of our synchronization using Mean Absolute Error (MAE), Root Mean Square Error (RMSE), and Mean Time Offset (MTO) between the ground truth event start times and the predicted start times from the synchronized data streams. The initial discrepancies between sensor streams averaged approximately 400 ms due to unsynchronized clocks and network delays. After applying our synchronization method, these discrepancies decreased to less than 50 ms, representing an 88\% reduction in temporal misalignment. The results are detailed in Table~\ref{table: metrics}. The significant reduction in synchronization error demonstrates the effectiveness of our method and the critical role of SASS's Multimodal Data Synchronization Services in achieving precise alignment across heterogeneous sensor modalities.

\subsection{Multicamera Detection}
Our multicamera detection application service greatly benefited from the Spatiotemporal Data Fusion Services offered by SASS. These services simplified the process of data acquisition, processing, and integration, allowing us to concentrate on high-level data fusion algorithms and application logic. As a result, we achieved a streamlined and resilient detection system, seamlessly embedded within the SASS framework.

The Spatiotemporal Data Fusion Services within SASS provided core functionalities that streamlined the development process. Fusion Workflow Engine, at the core of these services, orchestrates the fusion process by managing data flow between each component, enabling real-time fusion. It first yields time-aligned frames from cameras overlooking the area of interest. This is done using the location data and synchronized timestamps embedded in the metadata of the input data stream provided by Data Capture API. The flexibility of SASS in supporting multiple abstraction levels of fusion (early, intermediate, and late), allowed us to apply an intermediate fusion workflow by utilizing bounding box and confidence features obtain from raw sensor data processed by object an detection model.

\begin{table}[t]
\centering
\caption{Event Detection Accuracy: Camera (IMU)}
\label{table: accuracy}
\begin{tabular}{@{}lccc@{}} 
\toprule 
\textbf{Event Type} & \textbf{Recall} & \textbf{Precision} & \textbf{F1} \\
\midrule 
Gesture Event & 0.64 (0.61) & 1.00 (0.92) & \textbf{0.78 (0.73)} \\
Non-Event     & 1.00 (1.00) & 0.91 (0.98) & 0.95 (0.99) \\
\bottomrule 
\end{tabular}
\end{table}

\begin{table}[t]
\centering
\caption{Performance Metrics (Seconds) by Modality}
\label{table: metrics}
\begin{tabular}{@{}lccc@{}} % Alignment of the columns
\toprule % Top horizontal line
\textbf{Modality} & \textbf{MAE} & \textbf{RMSE} & \textbf{MTO} \\
\midrule % Middle horizontal line
Camera (IMU) & 0.054 (0.032) & 0.069 (0.039) & 0.033 (0.006) \\
Average & 0.04323 & 0.0545 & 0.0197 \\
\bottomrule % Bottom horizontal line
\end{tabular}
\end{table}

To transform the coordinates from each camera's perspective into a unified top-down view, we integrated our custom neural network, \textit{CoordinateTransformNet}, into the Fusion Algorithms Library provided by the Spatiotemporal Data Fusion Services. Instead of relying on the built-in inverse perspective calibration, this choice demonstrated the modular architecture and adaptability of the framework by allowing developers to integrate custom fusion algorithms tailored to specific application needs. Finally, the Fusion Workflow Engine handled integrating the custom processes into its overall workflow as part of a larger, adaptable fusion pipeline. By abstracting these computational complexities, SASS enabled the efficient deployment of our multicamera detection service.

\begin{figure}[t]
    \centering
    \includegraphics[width=\linewidth]{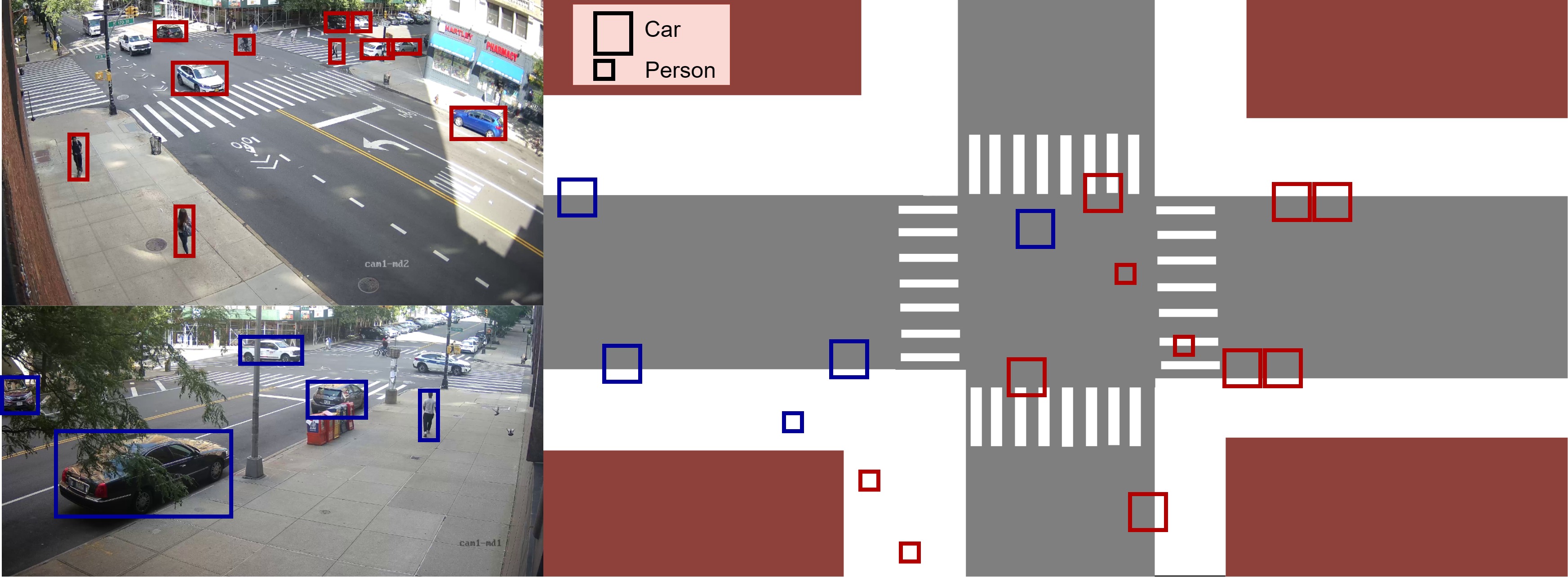}
    \caption{Detected bounding boxes from two perpendicular perspectives are integrated into the top-view perspective.}
    \label{fig:top-view}
\end{figure}

To further assess the impact of multi-camera integration on detection accuracy, we conducted experiments using a dataset of 900 images captured simultaneously from three cameras as illustrated in Figure~\ref{fig:mapping-points} (two street-level cameras on first and second floors and one high-altitude view camera on 12\textsuperscript{th}). To further enhance detection accuracy, we fine-tuned the object detection model using a manually annotated dataset recorded from the same high-altitude camera \cite{turkcan2024constellation}. As a result, the detection results from the 12\textsuperscript{th} floor camera can be considered a reliable ground truth. We performed object detection using the YOLOv9e model~\cite{wang2024yolov9} on the street-level cameras. The detection results were transformed using \textit{CoordinateTransformNet}, integrated within the Fusion Algorithms Library, as illustrated in Figure~\ref{fig:top-view}. The transformed detections were then compared against the ground truth from the high-altitude camera on temporally aligned frames.

\begin{figure}[t]
    \centering
    \includegraphics[width=.8\columnwidth]{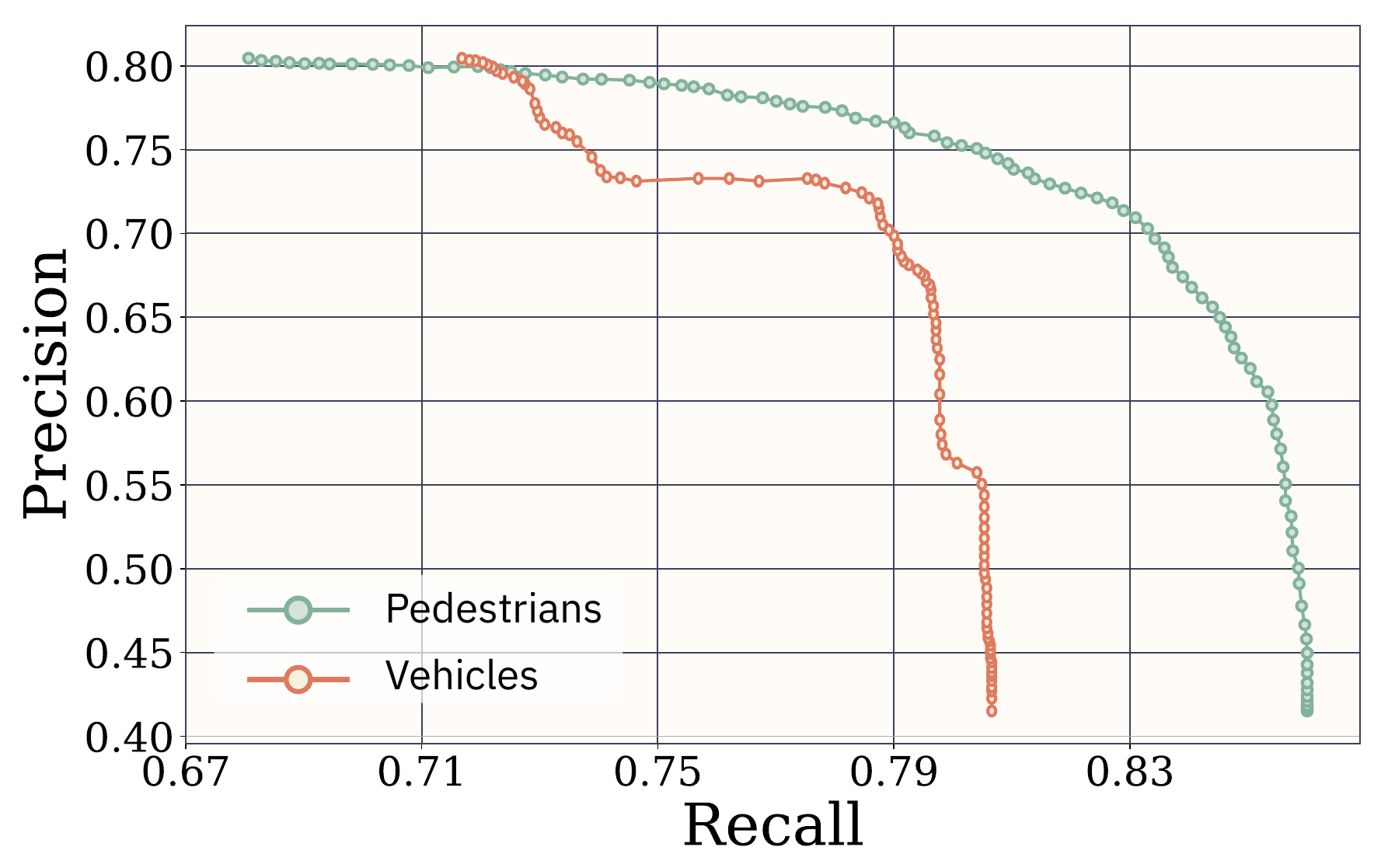}
    \caption{Precision-Recall Curves for Multi-Camera Integration, with the Duplicate Removal Threshold Reducing from 5.5\thinspace{m} to 0\thinspace{m} from Left to Right.}
    \label{fig:multicameraintegration}
\end{figure}

We present the precision, recall, and F1 score values for the pedestrian and vehicle classes separately in Table~\ref{table:combined-detection}. The results indicate that integrating the detections from both cameras enhances accuracy by over 10\% for both classes. We also analyzed the impact of varying the duplicate removal threshold on detection performance. Figure \ref{fig:multicameraintegration} shows the precision-recall curves for both classes as the threshold is reduced from 5.5 meters to 0 meters. For pedestrians, reducing the threshold led to a gradual decline in precision and a rise in recall, indicating fewer false positives but more false negatives. For vehicles, precision dropped more sharply, highlighting difficulties in accurately pinpointing vehicle centers. These results demonstrate the advantages of our method and the key role of SASS's Spatiotemporal Data Fusion Services in the seamless implementation of such multi-camera transformation models.

\subsection{Distributed Edge Processing}
The development and deployment of our real-time edge processing application service was significantly enhanced by the Edge Computing and Distributed Processing Services in SASS. These services streamlined task decomposition, computation distribution, and resource allocation, enabling us to maintain responsiveness and focus on high-level processing logic. This resulted in the processing of high-volume, computationally intensive tasks with minimal latency.

% \begin{table}[t]
% \centering
% \caption{Performance Metrics: Pedestrian and Vehicle Detection}
% \label{table:combined-detection}

% \begin{tabular}{@{}llccc@{}}
% \toprule
% \textbf{Class} & \textbf{Metric} & \textbf{1\textsuperscript{st} Cam} & \textbf{2\textsuperscript{nd} Cam} & \textbf{Multi-Cam} \\
% \midrule
% \multirow{3}{*}{Pedestrian} & Recall    & 0.609 & 0.813 & 0.790 \\
%                             & Precision & 0.769 & 0.635 & 0.766 \\
%                             & F1 Score  & 0.680 & 0.713 & \textbf{0.778} \\
% \midrule
% \multirow{3}{*}{Vehicle}    & Recall    & 0.681 & 0.558 & 0.720 \\
%                             & Precision & 0.510 & 0.684 & 0.617 \\
%                             & F1 Score  & 0.583 & 0.615 & \textbf{0.665} \\
% \bottomrule
% \end{tabular}
% \end{table}

\begin{table}[t]
\centering
\caption{Detection Metrics: Pedestrians (Vehicles)}
\label{table:combined-detection}

\begin{tabular}{@{}lccc@{}}
\toprule
\textbf{Metric} & \textbf{1\textsuperscript{st} Cam} & \textbf{2\textsuperscript{nd} Cam} & \textbf{Multi-Cam} \\
\midrule
Recall    & 0.609 (0.681) & 0.813 (0.558) & 0.790 (0.720) \\
Precision & 0.769 (0.510) & 0.635 (0.684) & 0.766 (0.617) \\
F1 Score  & 0.680 (0.583) & 0.713 (0.615) & \textbf{0.778 (0.665)} \\
\bottomrule
\end{tabular}
\end{table}

To support this application, SASS decomposed the primary task into several subtasks: mobile sensor capture, video frame acquisition and preprocessing, object detection, object tracking, trajectory calculation, 3D bounding box estimation, 2D and 3D pose estimation, and visualization. Each subtask was evaluated for computational load, with lighter tasks—such as frame capture/preprocessing, detection, tracking, and trajectory estimation—processed on edge medium nodes for low-latency results. Resource-intensive tasks, including 3D bounding box construction, 2D/3D pose estimation, and visualization, were offloaded to the Edge Computation Unit. This hierarchical distribution allows parallel subtask processing at independent rates, optimizing resource use and processing efficiency for downstream tasks. A key feature of this service, the scheduling algorithm, efficiently manages tasks on edge medium nodes by dynamically prioritizing them based on urgency and wait time. This prevents low-priority tasks from delaying high-priority, low-latency operations, optimizing throughput while preserving responsiveness across the entire subtasks. Additionally, SASS’s centralized resource monitor tracks node utilization in real time, enabling dynamic load management to reroute tasks as needed, preventing overloads, and ensuring consistent processing speeds.

\begin{figure*}[t]
    \centering
    \includegraphics[width=1\linewidth]{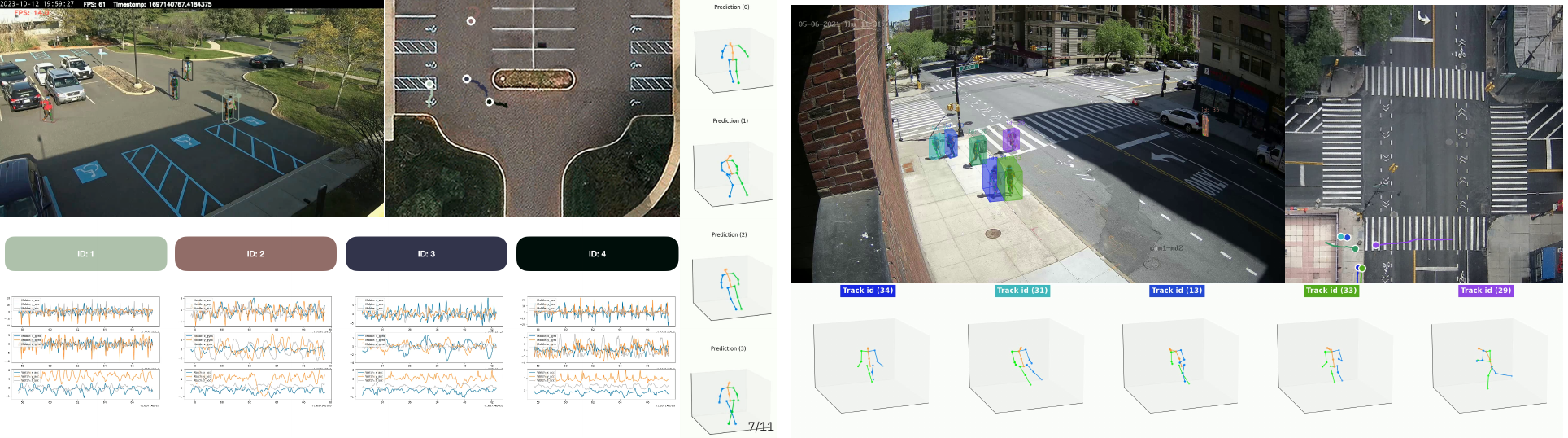}
    \caption{Analytics visualization dashboard showing live video feeds, IMU data streams, 3D bounding boxes, trajectories in both camera view and BEV, and 2D/3D pose estimations for pedestrians in urban and parking lot settings, processed with low-latency edge computing.}
    \label{fig:viz}
\end{figure*}

To further evaluate the performance of tasks within this module, the Edge Computing and Distributed Processing Services were tested for efficiency under demanding conditions. With an input video rate of 30 fps, the system achieves preprocessing, detection, tracking, and trajectory calculation results with a \textit{latency} of $\approx$30 milliseconds, enabling real-time operation of these subtasks. Computationally intensive stages, such as 3D bounding box construction and 2D/3D pose estimation, are seamlessly offloaded to the Edge Computation Unit, achieving a combined \textit{latency} of $\approx$45 milliseconds. Additionally, mobile sensor capture was handled directly on the Edge Computation Unit and used by the visualization dashboard, as these inputs originate from the application rather than edge sensors. The visualization task, which is resource-heavy due to 3D plotting, is processed at a latency of $\approx$200 milliseconds (Figure~\ref{fig:viz}). 

By processing subtasks independently and allowing downstream tasks to access data streams at customized output rates, SASS ensures flexible support for diverse application needs. The system \textit{throughput} for this pipeline reaches $\approx$57 tasks per second, facilitated by task segmentation and edge distribution. This is more than tenfold better than only 5 tasks per second without task decomposition, as the most resource-intensive task (visualization) operates at 5 fps. \textit{Task offloading} to the Edge Computation Unit constitutes about 47\% of total tasks, as these are more demanding processes.

We also evaluated the proposed scheduling algorithm across three key metrics. In our setup, we set initial priority value 0 and 1 for high- and low-priority tasks, and set $\alpha$ to 1. First, we measured the time taken by the scheduler to process, which revealed a low \textit{scheduling overhead} of average 0.07 milliseconds, ensuring minimal processing delay. We also measured the average time and variance from when each task enters the queue to when it is selected. The algorithm maintained an \textit{average latency} of 240 milliseconds and 4660 milliseconds and a \textit{wait time variance} of 390 milliseconds and 2670 milliseconds for high- and low-priority tasks, respectively. Lastly, we calculated the percentage of instances where lower-priority tasks were processed before higher-priority tasks due to the dynamic re-prioritization. The \textit{priority inversion rate} of 3\% underscores the algorithm's ability to sustain fair and responsive task processing.

\section{CONCLUSION} \label{conclusion}

This paper presented the Streetscape Application Services Stack (SASS), a novel framework that addresses the complexities of developing real-time, multimodal applications in urban environments. By providing modular services for multimodal data synchronization, spatiotemporal data fusion, and distributed edge computing, SASS abstracts the underlying challenges associated with integrating heterogeneous sensor modalities, varying spatial and temporal data, and low-latency edge processing.

Our contributions include the design of SASS as a programmable, scalable framework that offers uniform abstractions and controlled access to shared urban resources. We identified three essential requirements for urban applications—multimodal synchronization, data fusion, and low-latency edge processing—and implemented specialized services to support these needs. SASS enables developers to build and deploy complex, responsive applications through an SDK and APIs. We demonstrated SASS’s capabilities through three real-time application services: Multimodal Sensor Synchronization, Multicamera Detection, and Distributed Edge Processing. These services, supported by custom algorithms for event-based synchronization, multicamera fusion, and dynamic scheduling, showcase SASS’s adaptability for sophisticated urban analytics.

Furthermore, the promising performance metrics observed in our evaluation of these developed applications substantiate SASS’s ability to support the development of highly responsive and efficient urban applications. The Multimodal Data Synchronization service reduced temporal misalignment by 88\%, the Spatiotemporal Data Fusion service improved detection accuracy by over 10\%, and the Distributed Edge Computing service increased throughput tenfold, reaching 57 tasks per second. These results confirm that SASS’s modular design and efficient resource allocation can successfully facilitate complex, real-time streetscape applications, supporting future advancements in smart city solutions.

\begin{acks}
This work was supported by the National Science Foundation (NSF) and Center for Smart Streetscapes (CS3) under NSF Cooperative Agreement No. EEC-2133516, NSF Grant CNS-2148128, NSF Grant CNS-2038984, and corresponding support from the Federal Highway Administration (FHWA).
\end{acks}

\bibliographystyle{ACM-Reference-Format}
\bibliography{references}

% \appendix









\end{document}